\documentclass[11pt]{article}
\usepackage{amsmath,amssymb,color,epsfig,cite}
\usepackage{graphicx}
\usepackage{subfigure}
\usepackage{setspace}

\usepackage{amsfonts}

\newcommand{\hoch}[1]{$\, ^{#1}$}

\makeatletter
\@addtoreset{equation}{section}
\makeatother

\newcommand{\be}{\begin{equation}}
\newcommand{\ee}{\end{equation}}
\newcommand{\bea}{\setlength\arraycolsep{2pt} \begin{eqnarray}}
\newcommand{\eea}{\end{eqnarray}}
\newcommand{\nn}{\nonumber}

\def\0{{\sst{(0)}}}
\def\1{{\sst{(1)}}}
\def\2{{\sst{(2)}}}
\def\3{{\sst{(3)}}}
\def\4{{\sst{(4)}}}
\def\5{{\sst{(5)}}}
\def\6{{\sst{(6)}}}
\def\7{{\sst{(7)}}}
\def\8{{\sst{(8)}}}
\def\9{{\sst{(9)}}}

\def\sst#1{{\scriptscriptstyle #1}}

\begin{document}



\begin{center}
{\large {\bf Kerr-Sen Black Hole as Accelerator for Spinning Particles}}

\vspace{10pt}
Jincheng An, Jun Peng, Yan Liu, Xing-Hui Feng\hoch{*}

\vspace{15pt}

{\it Department of Physics, Beijing Normal University, Beijing 100875, China}

\vspace{30pt}

\underline{ABSTRACT}

\end{center}

It has been proved that arbitrarily high-energy collision between two particles can occur near the horizon of an extremal Kerr black hole as long as the energy $E$ and angular momentum $L$ of one particle satisfies a critical relation, which is called the BSW mechanism. Previous researchers mainly concentrate on geodesic motion of particles. In this paper, we will take spinning particle which won't move along a timelike geodesic into our consideration, hence, another parameter $s$ describing the particle's spin angular momentum was introduced. By employing the Mathisson-Papapetrou-Dixon equation describing the movement of spinning particle, we will explore whether a Kerr-Sen black hole which is slightly different from Kerr black hole can be used to accelerate a spinning particle to arbitrarily high energy. We found that when one of the two colliding particles satisfies a critical relation between the energy $E$ and the total angular momentum $J$, or has a critical spinning angular momentum $s_c$, a divergence of the center-of-mass energy $E_{cm}$ will be obtained.

\vfill {\footnotesize ~\\anjch@mail.bnu.edu.cn \\ jun\_peng@mail.bnu.edu.cn \\ yanliu@mail.bnu.edu.cn \\ *xhfengp@mail.bnu.edu.cn, corresponding author}

\pagebreak

\addtocontents{toc}{\protect\setcounter{tocdepth}{2}}


\newpage

\section{Introduction}

Ba\~nados, Silk, and West (BSW mechanism) first proved \cite{BSW} that an extremal Kerr black hole can be used as a particle accelerator in 2009, and the center-of-mass (CM) energy $E_{cm}$ of two test particles can be arbitrary high if the collision occurs near the extremal horizon and the energy $E$ and angular momentum $L$ of one particle satisfy a critical relation. Based on their pioneering discussion, the BSW mechanism has been extensively studied for multifarious black holes \cite{272}-\cite{1025}.

Most of the previous work on BSW mechanism focus on the acceleration of regular particles which are not spinning, thus such particles can be treated as point particles and move along timelike geodesics. However, a real particle is an extended body with self-interaction. Here  we will take spinning particles  which cannot be regarded as point particles into our consideration. It has been proved in\cite{Wald1972} and \cite{s1}\cite{s2}\cite{s3} that the movement of a spinning particle deviates from a geodesic due to the gravitational  interaction. The orbits of spinning particles has been computed based on the Mathisson-Papapetrou-Dixon (MPD) equation \cite{wave}\cite{z2}. Recently, some modification to the MPD equation was given in \cite{alx1}\cite{alx2}\cite{alx3}. The BSW mechanism for spinning particles in Kerr black hole has been studied in \cite{myg}, where the critical total angular momentum and critical spin is respectively
\begin{eqnarray}\label{l2e}
j_c=2,\quad s_c=1.
\end{eqnarray}

Also, the BSW mechanism for spinning particles in Kerr-Newman black hole has been studied in \cite{yuxiao}, where the  critical angular momentum and critical spin are given  respectively as
\be
j_c=\frac{1+a^2}{a},\quad s_c=\frac{1}{a}.
\ee
When we set the mass of the black hole $M=1$, for either Kerr or Kerr-Newman black hole, a physically acceptable value of $a$ must satisfy $0<a\leq1$, then we know that the critical spin for either Kerr or Kerr-Newman black hole,
\be\label{sc1}
s_c\geq1.
\ee
However, in a real world, the spin of an particle must be very small, i.e. $s\ll1$. Thus the critical spin is not applicable for Kerr or Kerr-Newman black hole. Also, spinning particle collision in Schwarzchild black hole background was discussed in \cite{zs}.

The no-hair theorem states that all black hole solutions of the Einstein-Maxwell theory can be completely characterized by only three parameters: mass, electric charge, and angular momentum. As we know, the Kerr-Newman metric is a solution for a rotating and charged black hole in the Einstein-Maxwell field equation. However, rotating and charged black hole solutions can also be found in other theories, such as string theory. Black holes in string theory can be coupled with other fields, such as the dilaton field, Yang-Mills field, and antisymmetric tensor gauge field. The Kerr-Sen black hole is a solution of the low-energy effective field theory for heterotic string theory \cite{Sen} and is also characterized by mass, electric charges, and angular momentum, which are similar to those of the Kerr-Newman black hole. However, the geometry of the Kerr-Sen black hole is different from that of the Kerr-Newman black hole.

Because string theory is the most promising candidate of unified theory, the expected rotating and charged black hole solution would be the Kerr-Sen solution rather than the Kerr-Newman one. In fact the BSW mechanism in the Kerr-Sen black bole background, i.e. the collision of point particles moving along timelike geodesic has been studied in \cite{275}, the critical relation for ultraenergetic collision given in it was also the same as that in \cite{BSW}. To further confirm such a critical relation, we will  study the collision of spinning particle that occurs in the horizon of an extremal Kerr-Sen black hole and check whether the relation can still hold for a divergent $E_{cm}$. Also, we want to explore if we can find a critical spinning angular momentum leading to an infinity of $E_{cm}$ like that in Kerr or Kerr-Newman black hole and check whether such a critical spin can be arbitrarily small, hence the requirement $s\ll1$ is satisfied.

The structure of this paper is as follows. At first, in section 2, we give a brief review on the Kerr-Sen black hole. Next, in section 3, we assumed the particles and the black hole have aligned/anti-aligned spin and the equatorial plane is chosen to be the orbital plane, then we will solve the equations of motion for the spinning particle. Then in section 4, we will analyse how the center-of-mass energy $E_{cm}$ of two spinning colliding particles behaves on the horizon of an extremal Kerr-Sen black hole and draw out a  critical total angular momentum and a critical spinning angular momentum, either of which will lead to a divergent $E_{cm}$. What's more, we also prove that the critical spin can meet the requirement $s\ll1$, hence it's applicable in Kerr-Sen black hole background. Finally, in section 5 we will further check the critical total angular momentum and the critical spin. We will testify that when the critical total angular momentum or critical spin is reached,  the collision of two spinning particles can indeed happen on the horizon of extremal Kerr-Sen black hole. In section 6, we will summarize all the main results we draw out in previous sections.

\section{Review of the Kerr-Sen black hole}\label{sec1}

We start this paper by giving a brief introduction to the Kerr-Sen black hole solution, which is described  by the 4-dimensional effective action of the heterotic string theory:
\begin{eqnarray}\label{action}
S=-\int d^{4}x\sqrt{-\mathcal{G}}e^{-\Phi}\bigg(-\mathcal{R}+\frac{1}{12}\mathcal{H}^{2}
-\mathcal{G}^{\mu\nu}\partial_{\mu}\Phi\partial_{\nu}\Phi+\frac{1}{8}\mathcal{F}^{2}\bigg),
\end{eqnarray}
where $\mathcal{R}$ is the scalar
curvature and $\Phi$ is the dilaton field,
$\mathcal{F}^{2}=\mathcal{F}_{\mu\nu}\mathcal{F}^{\mu\nu}$ with the field strength
$\mathcal{F}_{\mu\nu}=\partial_{\mu}\mathcal{A}_{\nu}-\partial_{\nu}\mathcal{A}_{\mu}$
corresponds to the Maxwell field
$\mathcal{A}_{\mu}$, and
\begin{eqnarray}
\mathcal{H}^{2}=\mathcal{H}_{\mu\nu\rho}\mathcal{H}^{\mu\nu\rho},
\end{eqnarray}
with $\mathcal{H}_{\mu\nu\rho}$ given by
\begin{eqnarray}
  \mathcal{H}_{\mu\nu\rho}&=&\partial_{\mu}\mathcal{B}_{\nu\rho}
                 +\partial_{\nu}\mathcal{B}_{\rho\mu}
                 +\partial_{\rho}\mathcal{B}_{\mu\nu}
                 -\frac{1}{4}\bigg(\mathcal{A}_{\mu}\mathcal{F}_{\nu\rho}
                 +\mathcal{A}_{\nu}\mathcal{F}_{\rho\mu}
                 +\mathcal{A}_{\rho}\mathcal{F}_{\mu\nu}\bigg),\label{Hmunurho}
\end{eqnarray}
where the last term in (\ref{Hmunurho}) is the gauge
Chern-Simons term. $\mathcal{G}_{\mu\nu}$ appeared in
(\ref{action}) are the covariant components of the metric in the
string frame, which are related to the Einstein metric by
$g_{\mu\nu}=e^{-\Phi}\mathcal{G}_{\mu\nu}$. The Einstein metric, the
non-vanishing components of $A_\mu$, $B_{\mu\nu}$ and the
dilaton field read \cite{Sen}:
\begin{eqnarray}
&&ds^{2}=-\bigg(\frac{\Delta-a^{2}\sin^{2}\theta}{\Sigma}\bigg)dt^{2}
         +\frac{\Sigma}{\Delta}dr^{2}
         -\frac{4\mu ar\cosh^{2}\alpha\sin^{2}\theta}{\Sigma}dtd\phi
         +\Sigma d\theta^{2}
         +\frac{\Xi\sin^{2}\theta}{\Sigma}d\phi^{2},\label{metric}~~~~~~\\
&&\mathcal{A}_{t}=\frac{\mu r\sinh 2\alpha}{\sqrt{2}\Sigma},
  \;\;\;\mathcal{A}_{\phi}=\frac{\mu a r\sinh 2\alpha\sin^{2}\theta}{\sqrt{2}\Sigma},\\
&&\mathcal{B}_{t\phi}=\frac{2a^{2}\mu r\sin^{2}\theta\sinh^{2}\alpha}{\Sigma},
  \;\Phi=-\frac{1}{2}\ln \frac{\Sigma}{r^{2}+a^{2}\cos^{2}\theta},
\end{eqnarray}
where the metric functions are given by
\begin{eqnarray}
&&\Delta=r^{2}-2\mu r+a^{2},\\
&&\Sigma=r^{2}+a^{2}\cos^{2}\theta+2\mu r\sinh^{2}\alpha,\\
&&\Xi=\bigg(r^{2}+2\mu r\sinh^{2}\alpha+a^{2}\bigg)^{2}
            - a^{2}\Delta\sin^{2}\theta.
\end{eqnarray}
When $\alpha=0$, the solution will return to the Kerr black hole.
The parameters $\mu$, $\alpha$ and $a$ are associated with the physical
mass $M$, the charge $Q$ and the angular momentum $J$ by
\begin{eqnarray}
  M=\frac{\mu}{2}(1+\cosh 2\alpha),\;\;
  Q=\frac{\mu}{\sqrt{2}}\sinh^{2}2\alpha,\;\;
  J=\frac{a\mu}{2}(1+\cosh 2\alpha).\label{equation}
\end{eqnarray}
Solving eq. (\ref{equation}), we can obtain
\begin{eqnarray}
 \sinh^{2}\alpha=\frac{Q^{2}}{2M^{2}-Q^{2}},\;\;
 \mu=M-\frac{Q^{2}}{2M}.\label{relation}
\end{eqnarray}
Then the parameters $\alpha$ and $\mu$ in the metric (\ref{metric})
can be eliminated. For a nonextremal black hole, there are two
horizons. They are both determined by $\Delta=0$ and are given by
\begin{eqnarray}
 r_{\pm}=M-\frac{Q^{2}}{2M}\pm\sqrt{\bigg(M-\frac{Q^{2}}{2M}\bigg)^{2}-a^{2}}.
 \label{horizons}
\end{eqnarray}
For the extremal black hole, it must be satisfied that
\begin{eqnarray}
 Q^{2}=2M(M-a) \quad \text{or} \quad \mu=a.
\end{eqnarray}
Setting $Q$ and $a$ to zero, respectively, we can obtain the maximum values for them. Thus,
we obtain the ranges for $a$ and $Q$, which are
\bea
&&0\le a \le M,\nn\\
&&0\le Q \le \sqrt2M.
\eea
Here, both the parameters $a$ and $Q$ are thought to be positive. For an extremal black hole,
the two horizons coincide with each other and the degenerate horizon locates at $r_0=a$.

By virtue of (\ref{relation}), we can get the inverse metric $g^{ab}$ with nonvanishing components
\begin{eqnarray}
&&g^{tt}=-\frac{\Xi\Sigma}{\Delta\Xi+a^{2}\sin^{2}\theta(4M^{2}r^{2}-\Xi)},\text{ }g^{t\phi}=-\frac{2aMr\Sigma}
      {\Delta\Xi+a^{2}\sin^{2}\theta(4M^{2}r^{2}-\Xi)},\nonumber\\
&&g^{rr}=\frac{\Delta}{\Sigma},\text{ }g^{\theta\theta}=\frac{1}{\Sigma},\text{ }g^{\phi\phi}=\frac{\Sigma(\Delta\csc^{2}\theta-a^{2})}
          {\Delta\Xi+a^{2}\sin^{2}\theta(4M^{2}r^{2}-\Xi)}.
\end{eqnarray}
And the metric functions can be expressed as
\begin{eqnarray}
&&\Delta=a^{2}+\frac{r}{M}\bigg(Q^{2}+Mr-2M^{2}\bigg),\\
&&\Sigma=\frac{r}{M}\bigg(Q^{2}+Mr\bigg)+a^{2}\cos^{2}\theta,\\
&&\Xi=\left[a^{2}+\frac{r}{M}(Q^{2}+Mr)\right]^{2}
          -a^{2}\Delta\sin^{2}\theta.
\end{eqnarray}
Here, we can explicitly see that the Kerr-Sen black hole is characterized
by three parameters, mass $M$, charge $Q$ and spin $a$. The Sen black hole solution will describe a Gibbon-Maeda
(GM) black hole with $a=0$ or describe a Kerr black hole with $Q=0$.

\section{Motion of spinning particle in Kerr-Sen black hole background}\label{sec2}

To describe the movement of spinning particles, we will employ the Mathisson-Papapetrou-Dixon equation \cite{Wald1972,myg}
\begin{eqnarray}
\frac{DP^a}{D\tau}=-\frac{1}{2}R^a{}_{bcd}v^bS^{cd},
\end{eqnarray}
\begin{eqnarray}
\frac{DS^{ab}}{D\tau}=P^av^b-P^bv^a,
\end{eqnarray}
where
\begin{eqnarray}
v^a=\Big(\frac{\partial}{\partial\tau}\Big)^a
\end{eqnarray}
is the tangent to the center of mass world line.
$\frac{D}{D\tau}$ is the covariant derivative along the world line, $P^a$ is spinning particle's 4-momentum  satisfying that
\begin{eqnarray}\label{n1s}
-m^2=P^aP_a,
\end{eqnarray}
$S^{ab}$ is the particle's spin tensor which is apparently antisymmetry $S^{ab}=-S^{ba}$ and has such a property that
\begin{eqnarray}\label{ans}
S^2=\frac{1}{2}S^{ab}S_{ab},
\end{eqnarray}
where $S$ is explained as the spinning angular momentum of the spinning particle.

There also exists the relation between $S^{ab}$ and $P^a$ as below
\begin{eqnarray}\label{sp}
S^{ab}P_b=0.
\end{eqnarray}
For a spinning particle, the conserved quantities associated with the $Killing $ vectors in its movement can be expressed as
\begin{eqnarray}\label{scq}
C_{\xi}=P^a\xi_{a}-\frac{1}{2}S^{ab}\bigtriangledown_a\xi_b.
\end{eqnarray}
To calculate the explicit form of $P^a$, let's rewrite the metric in tetrads as
\begin{eqnarray}
g_{ab}=\eta_{ij}e_a^{(i)}e_{b}^{(j)},
\end{eqnarray}
where the tetrad reads
\begin{eqnarray}\label{tetrads}
&&e_a^{(0)}=\sqrt{\frac{\Delta}{\Sigma}}\left(dt_a-a\sin^2\theta d\phi_a\right),\nonumber\\
&&e_a^{(1)}=\sqrt{\frac{\Sigma}{\Delta}}dr_a,\nonumber\\
&&e_a^{(2)}=\sqrt{\Sigma}d\theta_a,\nonumber\\
&&e_a^{(3)}=\frac{\sin{\theta}}{\sqrt{\Sigma}}\left[-adt_a+(r^2+2\mu r\sinh^2\alpha+a^2)d\phi_a\right].
\end{eqnarray}
Accordingly, the duality of these $e_a^{(i)}$ can be computed by
\begin{eqnarray}\label{eui}
e_{(i)}^a=\eta_{ij}e_b^{(j)}g^{ab}.
\end{eqnarray}
with
\begin{eqnarray}\label{invtetrads}
e^a_{(0)} &=& \frac{\sqrt{\Delta\Sigma}(\Xi-2a^2r\mu \cosh^2\alpha \sin^2\theta)}{\Delta\Xi+a^2\sin^2\theta(4r^2\mu^2\cosh^4\alpha-\Xi)}\Big(\frac{\partial}{\partial t}\Big)^a-\frac{\sqrt{\Delta\Sigma}a(\Delta-2r\mu \cosh^2\alpha+a^2 \sin^2\theta)}{\Delta\Xi+a^2\sin^2\theta(4r^2\mu^2\cosh^2\alpha-\Xi)}\Big(\frac{\partial}{\partial\phi}\Big)^a\nonumber\\
e^a_{(1)} &=& \sqrt{\frac{\Delta}{\Sigma}}\Big(\frac{\partial}{\partial r}\Big)^a,\nonumber\\
e^a_{(2)} &=& \frac{1}{\sqrt{\Sigma}}\Big(\frac{\partial}{\partial\theta}\Big)^a,\nonumber\\
e^a_{(3)} &=& -\frac{a(2a^2\mu r\cosh^2\alpha+2\mu r^3\cosh^2\alpha+4\mu^2r^2\sinh^2\alpha \cosh^2\alpha-\Xi)\sqrt{\Sigma}\sin\theta}{\Delta\Xi+a^2(4r^2\mu^2\cosh^4\alpha-\Xi)\sin\theta}\Big(\frac{\partial}{\partial t}\Big)^a\nonumber\\
&&+\frac{\sqrt{\Sigma}\csc\theta(\Delta[a^2+r(r+2\mu \sinh^2\alpha)]\csc^2\theta-a^2[a^2+r(r-2\mu)])}{a^2(4r^2\mu^2\cosh^2\alpha-\Xi)+\Delta\Xi \csc^2\theta}\Big(\frac{\partial}{\partial\phi}\Big)^a
\end{eqnarray}

For simplicity, let's take the particle's spin to be aligned with the black hole's spin, thus we can introduce a special spin vector $s^{(a)}$ which reads
\begin{eqnarray}
s^{(a)}=-\frac{1}{2m}\varepsilon^{(a)}{}_{(b)(c)(d)}\mu^{(b)}S^{(c)(d)},
\end{eqnarray}
equivalently,
\begin{eqnarray}
S^{(a)(b)}=m\varepsilon^{(a)(b)}{}_{(c)(d)}\mu^{(c)}s^{(d)},
\end{eqnarray}
where $\varepsilon_{(a)(b)(c)(d)}$ is a totally antisymmetric tensor with component $\varepsilon_{(0)(1)(2)(3)}=1$. As is argued in \cite{wave}, we can set the only non-zero component of $s^{(a)}$ as
\begin{eqnarray}
s^{(2)}=-s ,
\end{eqnarray}
where $s$ implies both the magnitude and direction of the particle's spin. The particle spin is parallel to the black hole spin for $s>0$, while it is antiparallel for $s<0$. Then we can get these non-vanishing components of the spin angular momentum in tetrad frame as below
\begin{eqnarray}\label{su}
&&S^{(0)(1)}=-ms\mu^{(3)},\nonumber\\
&&S^{(0)(3)}=ms\mu^{(1)},\nonumber\\
&&S^{(1)(3)}=ms\mu^{(0)}.
\end{eqnarray}
There are two Killing vector in Kerr-Sen black hole,
\begin{eqnarray}
\xi^a=\Big(\frac{\partial}{\partial t}\Big)^a,\quad\phi^a=\Big(\frac{\partial}{\partial \phi}\Big)^a.
\end{eqnarray}
Then with \eqref{tetrads}, we can get all the non-vanishing components of $\xi^a$, and $\phi^a$ as below
\begin{eqnarray}
\xi^{(0)}=\sqrt{\frac{\Delta}{\Sigma}},\quad\xi^{(3)}=-\frac{a\sin{\theta}}{\sqrt{\Sigma}},
\end{eqnarray}
\begin{eqnarray}
\phi^{(0)}=-a\sin{\theta}\sqrt{\frac{\Delta}{\Sigma}},\quad\phi^{(3)}=\frac{(r^2+2\mu r\sinh^2\alpha+a^2)\sin{\theta}}{\sqrt{\Sigma}},
\end{eqnarray}
Using \eqref{invtetrads}, we can obtain the non-vanishing components of covariant derivative of Killing vector in tetrad frame
\begin{eqnarray}
&&\nabla_{(0)}\xi_{(1)}=\frac{\mu \cosh^2\alpha(a^2\cos^2\theta-r^2)}{\Sigma^2},\nonumber\\
&&\nabla_{(0)}\phi_{(1)}=\frac{a\sin^2\theta[r(2(r+\mu \sinh^2\alpha)^2-r^2)+a^2\cos^2\theta(r-\mu)]}{\Sigma^2},\nonumber\\
&&\nabla_{(1)}\phi_{(3)}=-\frac{\sqrt{\Delta}\sin\theta(r+\mu \sinh^2\alpha)}{\Sigma^2},
\end{eqnarray}
For the convenience of following discussion, we define
\begin{eqnarray}\label{mu}
u^a=\frac{P^a}{m},
\end{eqnarray}
thus $u^au_a=-1$. Furthermore, we consider the case when the motion of the particle was confined on the equatorial plane $\theta=\frac{\pi}{2}$, which means $u^{(2)}=0$. Then \eqref{scq} can be rewritten as
\begin{eqnarray}
-E &=& -\xi^{(0)}u^{(0)}+\xi^{(3)}u^{(3)}+(\nabla_{(0)}\xi_{(1)})su^{(3)},\\
J &=& -\phi^{(0)}u^{(0)}+\phi^{(3)}u^{(3)}+(\nabla_{(0)}\xi_{(1)})su^{(3)}-(\nabla_{(1)}\phi_{(3)})su^{(0)},
\end{eqnarray}
with $\theta=\frac{\pi}{2}$, we have
\begin{eqnarray}
u^{(0)} &=& \frac{\Sigma_0 X}{\sqrt\Delta Z},\label{u0}\\
u^{(3)} &=& \frac{\Sigma_0^2 Y}{Z},
\end{eqnarray}
where
\begin{eqnarray}\label{xyz}
X &=& a^2E\Sigma_0^{3/2}+E\Sigma_0^{5/2}-aJ\Sigma_0^{3/2}+J\mu r^2\cosh^2\alpha s \nonumber\\
&&-aEr(\mu r+r^2+4\mu r\sinh^2\alpha+2\mu^2\sinh^4\alpha)s,\nonumber\\
Y &=& -aE\sqrt{\Sigma_0}+J\sqrt{\Sigma_0}+E(r+\mu \sinh^2\alpha)s,\nonumber\\
Z &=& \Sigma_0^3-\mu r^2\cosh^2\alpha(r+\mu \sinh^2\alpha)s^2,
\end{eqnarray}
$\Sigma_0$ is just the result of $\Sigma$ when $\theta=\frac{\pi}{2}$, i.e.
\begin{equation}
\Sigma_0 = r^2+2\mu r\sinh^2\alpha > 0.
\end{equation}
According to
\be
-(u^{(0)})^2+(u^{(1)})^2+(u^{(3)})^2 = -1,
\ee
we can obtain
\be
u^{(1)} = \sigma_r\frac{\sqrt{\cal R}}{\sqrt\Delta Z},\label{u1}
\ee
where $\sigma_r=\pm1$ and
\be
{\cal R} = \Sigma_0^2X^2-(\Sigma_0^4Y^2+ Z^2)\Delta.
\ee
With these result listed above, in the next section, we will explore the BSW mechanism of spinning particle in Kerr-Sen black hole background.

\section{ Spinning particles collision near horizon of extremal Kerr-Sen black hole}\label{sec3}

Here we consider that two uncharged spinning particles with the identical rest mass $m$ are at rest at infinity ($E=m$) and then they approach the black hole and collide near the horizon of an extremal $(\mu=a,r_0=a)$ Kerr-Sen black hole. For the sake of simplicity, we assume that the two particles have total angular momenta $j_1$ and $j_2$, spin $s_1$ and $s_2$ respectively. The center of mass energy is defined by
\be
E_{cm}^2 = -g_{ab}(P_1^a+P_2^a)(P_1^b+P_2^b),
\ee
considering \eqref{mu}, we have
\be
E_{cm}=\sqrt2m\sqrt{1-g_{ab}\mu_1^a\mu_2^b}.
\ee
For convenience, we only need to focus on the changeable part of $E_{cm}$, which is defined \cite{4D} as
\be\label{eff}
E_{eff} = -g_{ab}u_1^au_2^b = \frac{\Sigma_0^2X_1X_2-\sqrt{{\cal R}_1{\cal R}_2}}{\Delta Z_1Z_2}-\frac{\Sigma_0^4Y_1Y_2}{ Z_1Z_2}.
\ee
Because of the $\Delta$ factor in the denominator, it appears that $E_{eff}$ diverges at the horizon $r=r_0$. But this not true because, although not totally obvious, the numerator vanishes at that point as well. We can make an simple analysis.
Since $X$, $Y$, $Z$ are all finite, near the horizon $\Delta=0$, we have
\bea
\Sigma_0^2X_1X_2-\sqrt{{\cal R}_1{\cal R}_2}&=&\Sigma_0^2X_1X_2-\Sigma_0^2X_1X_2\sqrt{1-\frac{(\Sigma_0^4Y_1^2+ Z_1^2)\Delta}{\Sigma_0^2X_1}}\sqrt{1-\frac{(\Sigma_0^4Y_2^2+ Z_2^2)\Delta}{\Sigma_0^2X_2}}\nonumber\\
&=&\Sigma_0^2X_1X_2-\Sigma_0^2X_1X_2\left[1-\frac{(\Sigma_0^4Y_1^2+ Z_1^2)\Delta}{2\Sigma_0^2X_1}\right]\left[1-\frac{(\Sigma_0^4Y_2^2+ Z_2^2)\Delta}{2\Sigma_0^2X_2}\right]\nonumber\\
&\propto &\gamma_1\Delta+\gamma_2\Delta^2.
\eea
Then after taking it back into \eqref{eff}, we will find usually $E_{eff}$ is a finite term at the horizon.

Based on the rough discussion above, we know that $E_{eff}$ may not blow up casually, hence to find a possibility of an infinite $E_{eff}$, we need to expand the numerator of the first term in \eqref{eff} near the horizon, i.e.
\bea
\Sigma_0^2X_1X_2-\sqrt{{\cal R}_1{\cal R}_2}=\beta_0+\beta_1(r-a)+\beta_2(r-a)^2+\cdots
\eea
As expected, we find
\bea
 \beta_0=0,\quad\beta_1=0
\eea
and
\bea
\beta_2 &\propto& \frac{1}{(j_1-2a\cosh^2\alpha)(j_2-2a\cosh^2\alpha)}\times\nn\\
&&\frac{1}{(s_1\cosh^2\alpha-a\cosh^{\frac{3}{2}}2\alpha)(s_2\cosh^2\alpha-a\cosh^{\frac{3}{2}}2\alpha)}.
\eea
It should be pointed out that we have rescaled the total angular momentum $J$ by the mass $m$ of particle, i.e. $J\rightarrow jm$.
On the other hand
\be
Z_1Z_2|_{r=a} = (s_1^2\cosh^4\alpha-a^2\cosh^{3}2\alpha)(s_2^2\cosh^4\alpha-a^2\cosh^{3}2\alpha).
\ee
Finally the effective center of mass energy is given by
\bea
E_{eff} &=& \frac{K}{(j_1-2a\cosh^2\alpha)(j_2-2a\cosh^2\alpha)}\times\nn\\
&&\frac{1}{(s_1\cosh^2\alpha+a\cosh^{\frac{3}{2}}2\alpha)(s_2\cosh^2\alpha+a\cosh^{\frac{3}{2}}2\alpha)}\times\nn\\
&&\frac{1}{(s_1\cosh^2\alpha-a\cosh^{\frac{3}{2}}2\alpha)^2(s_2\cosh^2\alpha-a\cosh^{\frac{3}{2}}2\alpha)^2},
\eea
where $K$ is a very tedious  non-vanishing polynomial constructed by $(j_1,j_2,s_1,s_2,\alpha)$, hence we won't present it explicitly here.

We can see that when the angular momentum  of one of the particles satisfies a critical relation
\bea\label{cr}
j_c=2a\cosh^2\alpha,
\eea
or the spin of one of the particles satisfies a critical relation
\be
\quad s_c=\pm\frac{\cosh^{\frac{3}{2}}2\alpha}{\cosh^2\alpha}a
\ee
$E_{eff}$ will blow up.

According to \eqref{equation}, for extremal Kerr-Sen black hole ($\mu=a$), we have
\be
M = a\cosh^2\alpha.
\ee
So if we set $M=1$ conventionally, the critical angular momentum is just $j_c=2$, which is consistent with the result in \cite{BSW}. The critical spin becomes
\be
s_c = a^2\left(\frac{2}{a}-1\right)^{\frac{3}{2}}.
\ee
The range of $a$ is $0\le a\le1$. We can plot the critical spin $s_c$ as a function of $a$
\begin{center}
\includegraphics[height=4cm]{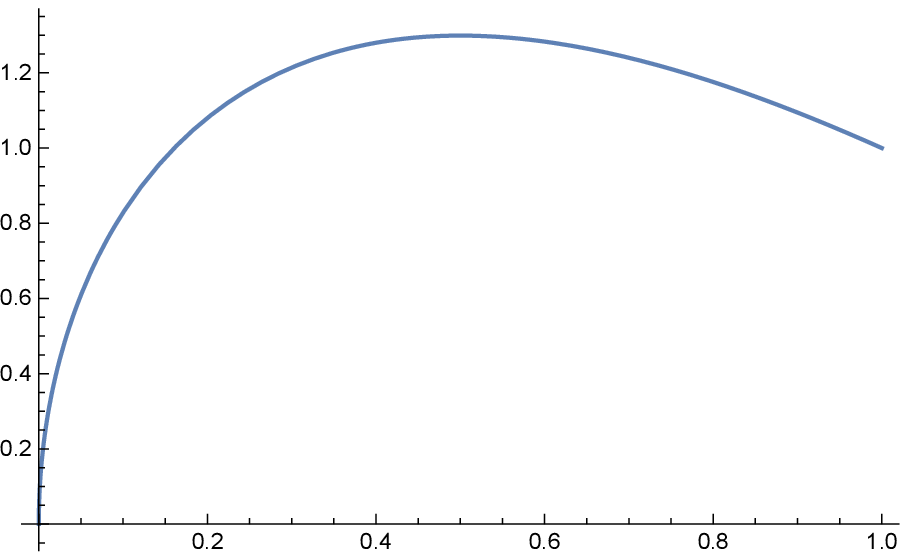}
\end{center}
The maximum of $s_c$ is $\frac{3\sqrt3}{4}$ when $a=\frac{1}{2}$. From above, we can know that different from the Kerr or Kerr-Newman case where the critical spin is unapplicable (see \eqref{sc1}), the critical spin in Kerr-Sen black hole background can be arbitrarily small as long as we choose a appropriate $a$. Hence the requirement $s\ll1$ is achieveable.

So far, we have draw out a critical relation that can result in a divergent $E_{cm}$. However, there is still something unavoidable for further discussion- with such a critical total angular momentum or critical spin, is a spinning particle really able to  collide another spinning particle on the horizon? The following section was devoted to this question.

\section{Can the collision be able to occur on the horizon?}\label{sec4}

To guarantee that the collision between two spinning particle can indeed occur near the horizon, we should impose such a requirement on the first approaching particle- when the first particle approach the horizon, it will move along the direction tangent to the horizon rather than getting through the horizon and falling into the black hole. Only in this way, when the second particle approach the horizon, can it be able to collide with the first particle. Hence, we need to check with the critical relation \eqref{cr}, whether a spinning can move along the direction tangent to the horizon.

The radial equation of motion is
\be
\frac{\dot r^2}{2}+V_{eff} = 0,
\ee
where $V_{eff}$ is the effective potential and it is given by
\be
V_{eff} = -\frac{\dot r^2}{2} = -\frac{1}{2}(e^1_{(1)}u^{(1)})^2 = -\frac{{\cal R}}{2\Sigma Z^2}.
\ee
The condition under which a spinning particle can move tangent the horizon is that
\be
V_{eff}\mid_{r=r_0} = 0,\quad \partial_rV_{eff}\mid_{r=r_0} = 0,
\ee
which is equivalent to
\be\label{rdr}
{\cal R}\mid_{r=r_0} = 0,\quad \partial_r{\cal R}\mid_{r=r_0} = 0.
\ee

\bigskip
\noindent{\bf Critical total angular momentum}:
\medskip

Noticing when the black hole is extremal, in ${\cal R}$, there is a $\Delta=(r-r_0)^2$ in the second term, so we only need to concentrate on $X$ in the first term.
For extremal Kerr-Sen black hole, when $j$ take the critical angular momentum $j_c=2a\cosh^2\alpha$, we have
\be
X = (r-r_0)[\Sigma_0^{3/2}(r+a(1+2\sinh^2\alpha))-ras(r-2a\sinh^4\alpha)],
\ee
Then we know that
\be
{\cal R} = (r-r_0)^2\tilde R,
\ee
where
\bea
\tilde R &=& \Sigma_0^2[\Sigma_0^{3/2}(r+a(1+2\sinh^2\alpha))-ras(r-2a\sinh^4\alpha)]^2\nn\\
&&-\Sigma_0^4[-a\sqrt\Sigma_0+2a\cosh^2\alpha\sqrt\Sigma_0+(r+a\sinh^2\alpha)s]^2\nn\\
&&-[\Sigma_0^3-ar^2\cosh^2\alpha(r+a\sinh^2\alpha)s^2]^2.
\eea
thus the requirement \eqref{rdr} is satisfied. But this is not the end of the story. As expected, the effective potential $V_{eff}$ approaches 0 at infinity. We must require
\be
V_{eff}\le0\quad\text{for any}\quad r\ge r_0,
\ee
for the particle falling freely from rest at infinity can reach the horizon. Because of $\Sigma_0>0$, we get
\be
\tilde R\ge0\quad\text{for any}\quad r\ge r_0.
\ee
This condition can also be obtained by the radial component $u^1$ of 4-velocity given by \eqref{u1}. There for one particle can reach the horizon, we must have ${\cal R}>0$, which results also to $\tilde R\ge0$.

When the spin of particle is zero, i.e. $s=0$, we have
\be
\tilde R = 2ar\Sigma_0^5\cosh^2\alpha,
\ee
which is always positive.

\bigskip
\noindent{\bf Critical spin angular momentum}:
\medskip

For extremal Kerr-Sen black hole, when $s$ take the critical spin $s_c=\frac{\cosh^{\frac{3}{2}}2\alpha}{\cosh^2\alpha}a$, we can show that
\be
X = c_1(r-a)+c_2(r-a)^2+\cdots,
\ee
so the requirement \eqref{rdr} is satisfied. But for another branch $s_c=-\frac{\cosh^{\frac{3}{2}}2\alpha}{\cosh^2\alpha}a$, we see that
\be
X = X_0+d_1(r-a)+d_2(r-a)^2+\cdots,
\ee
which doesn't satisfied the requirement \eqref{rdr}. Thus, only the critical spin $s_c=\frac{\cosh^{\frac{3}{2}}2\alpha}{\cosh^2\alpha}a$ can help the particle move tangent to the horizon, hence, guaranteeing the occurrence of the collision. In other worlds, the spin of particle must be parallel to the rotation of black hole.

In the same way, for the particle falling freely from rest at infinity can reach the horizon, we must require
\be
{\cal R}\ge0\quad\text{for any}\quad r\ge r_0.
\ee

\section{Conclusion}

In this paper, we have investigated the collision of two uncharged spinning particles (which could be thought to be the cold dark matter particles) falling freely from rest at infinity in Kerr-Sen spacetime. The Kerr-Sen black hole is a solution of four dimensional low energy effective supergravity of the Heterotic string theory. It is characterized by three parameters, mass $M$, charge $Q$ and spin $a$. When the rotation $a$ vanishes, it describes a Gibbon-Maeda black hole. When the charge $Q$ vanishes, it describes a Kerr black hole. Naively, one may think that Kerr-Sen black hole have no big difference with Kerr-Newmann black hole. But our results show that  they are quite different. For extremal Kerr-Sen black hole, it was found that the CM energy of two spinning particles can be divergent when the total angular momentum $j$ of one of the two particles reaches its critical value $j_c = 2$, which is the same as the case of extremal Kerr black hole. Furthermore, if the spin of one of the particles satisfied $s_c =a^2\left(\frac{2}{a}-1\right)^{3/2}$, the CM energy might also be divergent. Different from the Kerr or Kerr-Newman case where the critical spin is unapplicable (see \eqref{sc1}), the critical spin in Kerr-Sen black hole background can be arbitrarily small as long as we choose a appropriate $a$. We  also discussed the condition under which the collision between two particles can take place in fact near the horizon of extremal Kerr-Sen black hole.

At present,  we performed our calculations without considering the back reaction effect of the accelerated particle pair on the background geometry of the Kerr-Sen black hole. Due to the energy  carried by the infalling particles, a new and larger horizon is formed before the particles reach the horizon of the original black hole. This in fact implies, in particular, that the optimal collision between the two particles cannot take place exactly at the horizon of the original black hole \cite{Hod:2016kzj}. This issue deserves further research in the future.

\section*{Acknowledgments}

J.A. is supported in part by NSFC grants No.~11375026 and No.~11235003. J.P. and Y.L. are supported in part by NSFC grants No.~11235003. X-H.F. is supported in part by NSFC grants No.~11475024, No.~11175269 and No.~11235003.


\begin{thebibliography}{3}

\bibitem{BSW}
  M.~Banados, J.~Silk and S.~M.~West,
  ``Kerr Black Holes as Particle Accelerators to Arbitrarily High Energy,''
  Phys.\ Rev.\ Lett.\  {\bf 103}, 111102 (2009)
  [arXiv:0909.0169 [hep-ph]].

\bibitem{272}
  O.~B.~Zaslavskii,
  ``Acceleration of particles as universal property of rotating black holes,''
  Phys.\ Rev.\ D {\bf 82}, 083004 (2010)
  [arXiv:1007.3678 [gr-qc]].

\bibitem{273}
  O.~B.~Zaslavskii,
  ``Acceleration of particles by black holes: general explanation,''
  Class.\ Quant.\ Grav.\  {\bf 28}, 105010 (2011)
  [arXiv:1011.0167 [gr-qc]].

\bibitem{274}
  S.~W.~Wei, Y.~X.~Liu, H.~Guo and C.~E.~Fu,
  ``Charged spinning black holes as Particle Accelerators,''
  Phys.\ Rev.\ D {\bf 82}, 103005 (2010)
  [arXiv:1006.1056 [hep-th]].

\bibitem{275}
  S.~W.~Wei, Y.~X.~Liu, H.~T.~Li and F.~W.~Chen,
  ``Particle Collisions on Stringy Black Hole Background,''
  JHEP {\bf 1012}, 066 (2010)
  [arXiv:1007.4333 [hep-th]].

\bibitem{276}
  M.~Kimura, K.~i.~Nakao and H.~Tagoshi,
  ``Acceleration of colliding shells around a black hole: Validity of the test particle approximation in the Banados-Silk-West process,''
  Phys.\ Rev.\ D {\bf 83}, 044013 (2011)
  [arXiv:1010.5438 [gr-qc]].

\bibitem{277}
  T.~Harada and M.~Kimura,
  ``Collision of two general geodesic particles around a Kerr black hole,''
  Phys.\ Rev.\ D {\bf 83}, 084041 (2011)
  [arXiv:1102.3316 [gr-qc]].

\bibitem{278}
  C.~Zhong and S.~Gao,
  ``Particle collisions near the cosmological horizon of a Reissner-Nordstr\'om de Sitter black hole,''
  JETP Lett.\  {\bf 94}, 589 (2011)
  [arXiv:1109.0772 [hep-th]].
\bibitem{1025}A. Galajinsky,
``Particle collisions on near horizon extremal Kerr background,''  Phys. Rev. D {\bf88,}  027505(2013).
\bibitem{4D}
  S.~Gao and C.~Zhong,
  ``Non-extremal Kerr black holes as particle accelerators,''
  Phys.\ Rev.\ D {\bf 84}, 044006 (2011)
  [arXiv:1106.2852 [gr-qc]].

\bibitem{Wald1972}
  R.~M.~Wald,
  ``Gravitational spin interaction,''
  Phys.\ Rev.\ D {\bf 6}, 406 (1972).

\bibitem{s1}
  M.~Mathisson,
  ``Neue mechanik materieller systemes,''
  Acta Phys.\ Polon.\  {\bf 6}, 163 (1937).

\bibitem{s2}
  A.~Papapetrou,
  ``Spinning test particles in general relativity. 1.,''
  Proc.\ Roy.\ Soc.\ Lond.\ A {\bf 209}, 248 (1951).

\bibitem{s3}
  W.~G.~Dixon,
  ``Dynamics of extended bodies in general relativity. I. Momentum and angular momentum,''
  Proc.\ Roy.\ Soc.\ Lond.\ A {\bf 314}, 499 (1970).

\bibitem{alx1}
  A.~A.~Deriglazov and W.~G.~Ram\'irez,
  ``Mathisson-Papapetrou-Tulczyjew-Dixon equations in ultra-relativistic regime and gravimagnetic moment,''
  Int.\ J.\ Mod.\ Phys.\ D {\bf 26}, no. 06, 1750047 (2016).

  \bibitem{alx2}
  A.~A.~Deriglazov and W.~G.~Ram\'irez,
  ``Ultrarelativistic Spinning Particle and a Rotating Body in External Fields,''
  Adv.\ High Energy Phys.\  {\bf 2016}, 1376016 (2016).
  \bibitem{alx3}
  W.~G.~Ram\'irez and A.~A.~Deriglazov,
  ``Relativistic effects due to gravimagnetic moment of a rotating body,''
  arXiv:1709.06894 [gr-qc].
\bibitem{wave}
  M.~Saijo, K.~i.~Maeda, M.~Shibata and Y.~Mino,
  Phys.\ Rev.\ D {\bf 58}, 064005 (1998).


\bibitem{z2}
  P.~I.~Jefremov, O.~Y.~Tsupko and G.~S.~Bisnovatyi-Kogan,
  ``Innermost stable circular orbits of spinning test particles in Schwarzschild and Kerr space-times,''
  Phys.\ Rev.\ D {\bf 91}, no. 12, 124030 (2015)
  [arXiv:1503.07060 [gr-qc]].

\bibitem{myg}
  M.~Guo and S.~Gao,
  ``Kerr black holes as accelerators of spinning test particles,''
  Phys.\ Rev.\ D {\bf 93}, no. 8, 084025 (2016)
  [arXiv:1602.08679 [gr-qc]].

\bibitem{yuxiao}
  Y.~P.~Zhang, B.~M.~Gu, S.~W.~Wei, J.~Yang and Y.~X.~Liu,
  ``Charged spinning black holes as accelerators of spinning particles,''
  Phys.\ Rev.\ D {\bf 94}, no. 12, 124017 (2016)
  [arXiv:1608.08705 [gr-qc]].
 \bibitem{zs}
  O.~B.~Zaslavskii,
  ``Schwarzschild black hole as particle accelerator of spinning particles,''
  EPL {\bf 114}, no. 3, 30003 (2016).
\bibitem{Sen}
  A.~Sen,
  ``Rotating charged black hole solution in heterotic string theory,''
  Phys.\ Rev.\ Lett.\  {\bf 69}, 1006 (1992)
  [hep-th/9204046].

\bibitem{Hod:2016kzj}
  S.~Hod,
  ``Upper bound on the center-of-mass energy of the collisional Penrose process,''
  Phys.\ Lett.\ B {\bf 759}, 593 (2016)
  [arXiv:1609.06717 [gr-qc]].

\end{thebibliography}
\end{document}